\documentclass[11pt]{article}
\usepackage[dvips]{graphicx}
\usepackage{amssymb}

\def\dg{\mbox{$^\circ$}}		%degree sign
\def\etal{et al.}
\def\hMpc{h^{-1}{\rm Mpc}}

\newcommand{\area}{4278}
\newcommand{\pcarea}{40}

\newcommand{\plate}{621}
\newcommand{\galspec}{264,995}
\newcommand{\qsospec}{37,612}
\newcommand{\starspec}{50,023}
\newcommand{\nspec}{350,000}

\begin{document}

\title{The Sloan Digital Sky Survey}

\author{Jon Loveday, for the SDSS collaboration\\
Astronomy Centre, University of Sussex, Falmer, Brighton, BN1 9QJ,
England}

\maketitle

\begin{abstract}
The Sloan Digital Sky Survey (SDSS) is making a multi-colour, three dimensional
map of the nearby Universe.
The survey is in two parts.
The first part is imaging one quarter of the sky in five colours 
from the near ultraviolet to the near infrared.
In this imaging survey we expect to detect around 50 million galaxies to a 
magnitude limit $g \sim 23$.
The second part of the survey, taking place concurrently with the imaging,
is obtaining spectra for up to 1 million galaxies and 100,000 quasars.
From these spectra we obtain redshifts and hence distances, in order to map
out the three-dimensional distribution of galaxies and quasars in the Universe.
These observations will be used to constrain models of cosmology and of 
galaxy formation and evolution.

This article describes the goals and methods used by the SDSS, the current 
status of the survey, and highlights some exciting discoveries made from
data obtained in the first two years of survey operations.
\end{abstract}

\section{Introduction}

What is the Universe made of?
How did the Universe begin?
How will it end?

These are some of the fundamental questions which can be addressed by
studying the large scale distribution of galaxies in the Universe.
It is widely believed that the galaxies we see today formed at the sites of
tiny ($\sim$ 1 part in $10^5$) density fluctuations in the early Universe.
The form of these density fluctuations (which, if Gaussian in nature,
may be fully described by their power spectrum) are predicted by cosmological
models, and depend on such parameters as the mean matter density $\Omega_m$,
the fraction of baryonic matter $\Omega_b/\Omega_m$ and any contribution to
the cosmological density from vacuum energy, also known as the 
cosmological constant $\Omega_\Lambda$.
On large scales, the clustering of galaxies 
can be predicted from the primordial density fluctuations using linear
perturbation theory.
By measuring this large-scale clustering, we can thus 
obtain important constraints on cosmological models.
By studying the intrinsic properties of the galaxies themselves, such as
luminosity, colour and morphology, we can test theories for how galaxies
are born and evolve.

In order to measure the clustering of galaxies reliably, it is
important to use a systematic and well-defined catalogue.
Systematic surveys date back to that of Messier, published in three parts
in the 1770s and 1780s, although it was not realized at the time that
some of the nebulae catalogued by Messier 
were other galaxies outside our own Milky Way.
It was only in 1923, by a careful measurement of the distance to the 
Andromeda Nebula (M31 in Messier's catalogue), 
that Edwin Hubble proved definitively that M31 was a 
large galaxy separate from our own.
Hubble later discovered the expansion of the Universe, and found that
the recession velocity of a galaxy is in direct proportion to its
distance from us, the Hubble law.
Since then, a number of galaxy surveys have been published,
starting with that of Shapley and Ames in 1932 \cite{sa1932}, 
and including more recently
the APM Galaxy Survey \cite{msel90}, which contains positions 
and magnitudes for about three million galaxies.
Most of these surveys are based on photographic plates, and
there is concern that such surveys could be missing a 
substantial fraction of low surface-brightness galaxies, eg.~\cite{disney76}.
There is also apprehension that uncertainties in the photometric calibration
of these surveys could lead to spurious measurement of galaxy clustering
on large scales \cite{fhs92,mes96}.

Smaller surveys have been made using charge coupled device (CCD) detectors.
These solid state devices, unlike
photographic plates, have a linear response to light and $\sim 50$ times higher
quantum efficiency, but the limited size of
these detectors has before now precluded the construction of wide-area galaxy 
surveys.

In order to map out the three-dimensional distribution of galaxies,
as opposed to just their two-dimensional projection on the celestial sphere,
one needs the distance to each galaxy.
This may be obtained by measuring the spectrum of light emitted by a galaxy.
The Doppler shift in features towards the red end of the spectrum
(the redshift) may be used to infer a galaxy's recession velocity and hence
its distance from the Hubble relation.
Until recently, galaxy spectra were painstakingly measured one-by-one, 
and it is only in the last few years that optical fibre multiplexing has
been used to measure redshifts for many thousands of galaxies.

The largest redshift survey to date is the nearly-completed Two Degree Field 
(2dF) Galaxy Redshift Survey carried out on the Anglo-Australian Telescope
\cite{colless2001}.
While containing redshifts for more than 200,000 galaxies, the 2dF survey is
based on the photographic APM Galaxy Survey, with the potential problems
mentioned above.

The Sloan Digital Sky Survey (SDSS) collaboration was therefore formed in 
1988 with the aim
of constructing a definitive map of the local universe, incorporating
CCD imaging in several passbands over a large area of sky, and measurement of
redshifts for around one million galaxies.
In order to complete such an ambitious project over a reasonable timescale,
it was decided to build a dedicated 2.5-metre telescope
equipped with a large CCD array imaging camera and multi-fibre spectrographs.
The survey itself began in April 2000, and observations are scheduled to
finish in June 2005.
In this article I review some important aspects of the survey,
including an overview of survey operations (\S\ref{sec:overview}), 
a description of the preliminary public data release (\S\ref{sec:edr}), 
and a selection of some early science results (\S\ref{sec:science}).

\section{Survey Overview}
\label{sec:overview}

\subsection{Goals}

The basic goal of the survey is to make a definitive map of the local Universe,
which can then be used to constrain cosmological models and models
for the formation and evolution of galaxies.
This map will consist of 5-colour imaging to a $g$-band limiting magnitude
$m \approx 23$
over a contiguous area of $\pi$ steradians (10,000 square degrees) 
in the northern sky and three, non-contiguous stripes with total area 
740 square degrees in the southern sky.
(Magnitudes measure flux on a logarithmic scale, where smaller
magnitudes correspond to larger flux.
The magnitude difference between two stars of flux $f_1$ and $f_2$ is
defined to be $m_1 - m_2 = -2.5 \lg(f_1/f_2)$.
The brightest stars have magnitude $m \approx 1$, the faintest stars 
visible to the naked eye have $m \approx 6$, ie. 100 times fainter.
A magnitude $m \approx 23$ thus corresponds to an observed flux which is
roughly six million times fainter than that of the dimmest naked-eye stars.)
From this imaging data we can map out the two-dimensional distribution
of galaxies and quasars as projected on the celestial sphere.
Distances to a subset of these objects will be determined by observing
spectra of roughly one million galaxies and 100,000 quasars.
The spectrum of a galaxy or quasar enables one to measure its recession
velocity $v$ from the Doppler redshift of spectral features.
Distances $d$ may then be estimated from the Hubble law: $v = H_0 d$, 
where $H_0 \approx 70$ km/s/Mpc is the {\em Hubble parameter}.
In this way, we can map out the three-dimensional distribution of objects 
in space.
(Astronomers often measure extragalactic distances in units of
Mpc where 1 Mpc = $10^6$ parsecs, and 1 parsec $\approx 3.09 \times 10^{16}$m.
Uncertainty in the value of $H_0$ leads to a corresponding uncertainty
in distances derived from the Hubble relation.
To reflect this, distances are often written in units of $\hMpc$,
where $H_0 = 100 h$ km/s/Mpc.)

\subsection{Hardware and operations}

The SDSS consists of two concurrent surveys, 
one photometric and one spectroscopic.
To complete a digital survey over a large fraction of the sky within a 
reasonable timescale, it is necessary to conduct wide-field imaging and 
multi-object spectroscopy.  To meet this
need, a wide-field telescope, imaging camera and multi-fibre spectrographs
were designed and built specifically
for this purpose, which I describe very briefly below.

The survey hardware comprises the 2.5-metre survey telescope, 
a 0.5-metre photometric telescope (called the monitor telescope in
its previous incarnation),
a state-of-the-art imaging camera \cite{gunn98} that observes 
near-simultaneously
in five passbands covering the near-ultraviolet to near-infrared,
and a pair of dual beam 
spectrographs, each capable of observing 320 fibre-fed spectra.
The site is also equipped with a 10-micron all-sky camera \cite{hls94},
which provides rapid warning of any cloud cover.
The data are reduced by a series of automated pipelines and the
resulting data products stored in an object-oriented database.

The main survey telescope is of modified Ritchey-Chr\'etien design
\cite{wmgk98}, with a primary aperture of 2.5m and a focal ratio
of f/5 to produce a flat field of 3\dg\ with a plate scale of 16.51
arcsec/mm.  It is situated at Apache Point Observatory, near Sunspot, New
Mexico, at a height of 2,800m.  The telescope is housed in an enclosure
which rolls off for observing, and is encased in a co-rotating baffle which
protects it from wind disturbances and stray light.
This unique design allows the telescope to remain free of dome-induced seeing.
Photographs of the site and telescope can be found at {\tt http://www.sdss.org/}.
A technical overview of the survey has been given by York et al. 
\cite{york2000} and further details are available online in the 
{\em SDSS Project Book} at {\tt http://astro.princeton.edu/PBOOK/welcome.htm}.

\subsection{Imaging Survey}

The photometric imaging survey will produce a database of roughly 10$^8$ 
galaxies and 10$^8$ stellar objects, with accurate ($\leq$ 0.10 arcsec) 
astrometry, 5-colour photometry, 
and object classification parameters.  This database
will become a public archive.

\begin{figure}
\includegraphics[width=\textwidth]{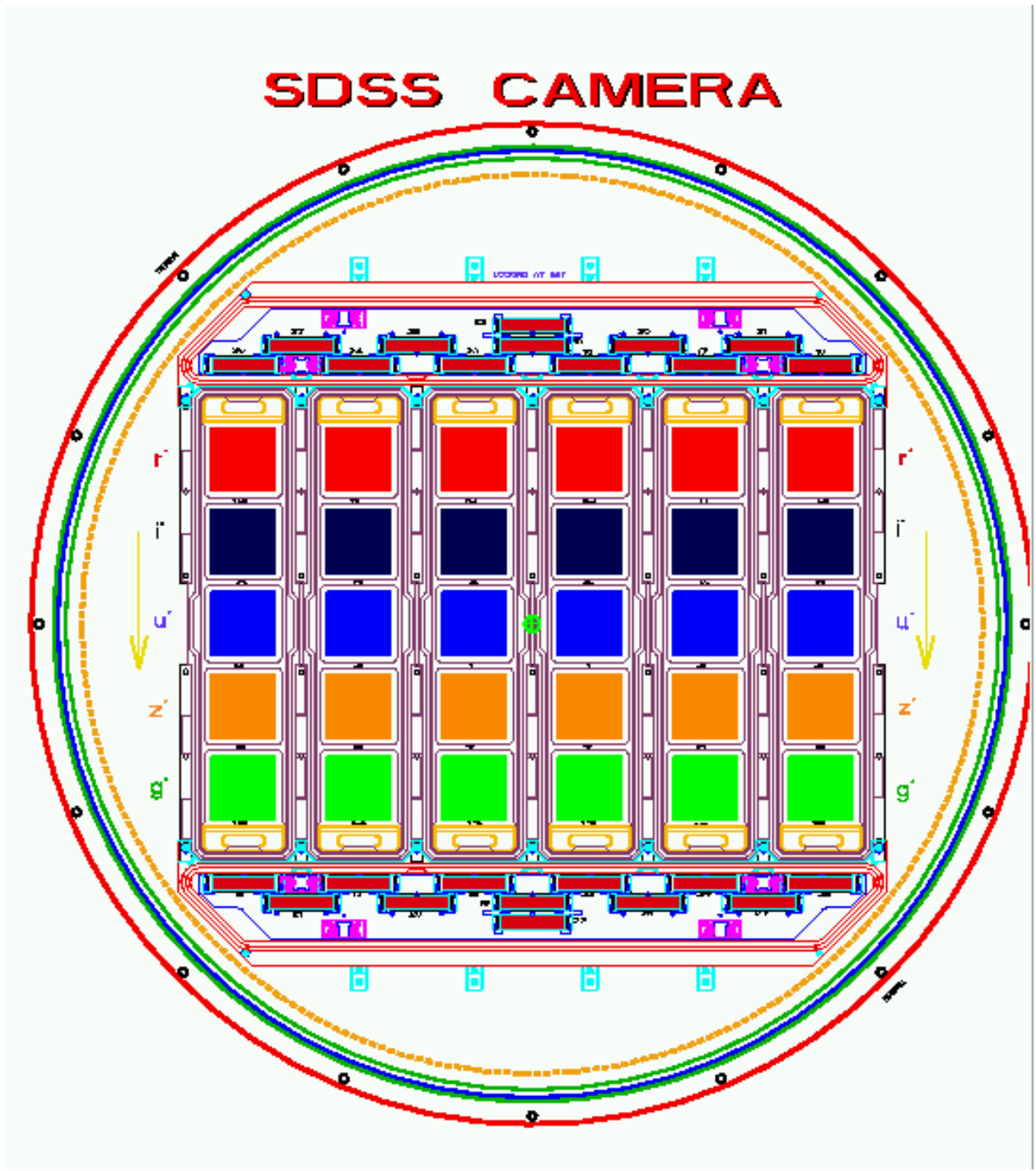}
\caption{Front view of the imaging camera assembly.
The $ugriz$ imaging CCDs are colour-coded blue, green, red, black and
orange respectively.
The astrometric CCDs are shown as narrow red rectangles.
The arrows indicate the direction of motion of astronomical sources
across the imaging array, which subtends 2.3\dg\ on the sky.}
\label{fig:camera}
\end{figure}

The imaging camera \cite{gunn98} (Figure~\ref{fig:camera})
consists of 54 CCDs in eight dewars and spans $2.3\dg$ on the sky.  
Thirty of these CCDs
are the main imaging/photometric devices, each a SITe (Scientific
Imaging Technologies, formerly Tektronix) device with $2048 \times 2048\ 24\mu$
pixels.  They are arranged in six dewars aligned with the scan direction
and holding 5 CCDs each, one CCD for
each filter bandpass.  The camera
operates in TDI (Time Delay and Integrate), or scanning, mode for which
the telescope
is driven at a rate synchronous with the charge transfer rate of
the CCDs.  Objects on the sky drift down the CCD array so
that nearly simultaneous 5-colour photometry is obtained.
The effective integration time,
{\it i.e.} the time any part of the sky spends on each detector, is 55 seconds
at the chosen (sidereal) scanning rate, which results in a limiting
magnitude of $g \sim 23$.

\begin{figure}
\includegraphics[width=\textwidth]{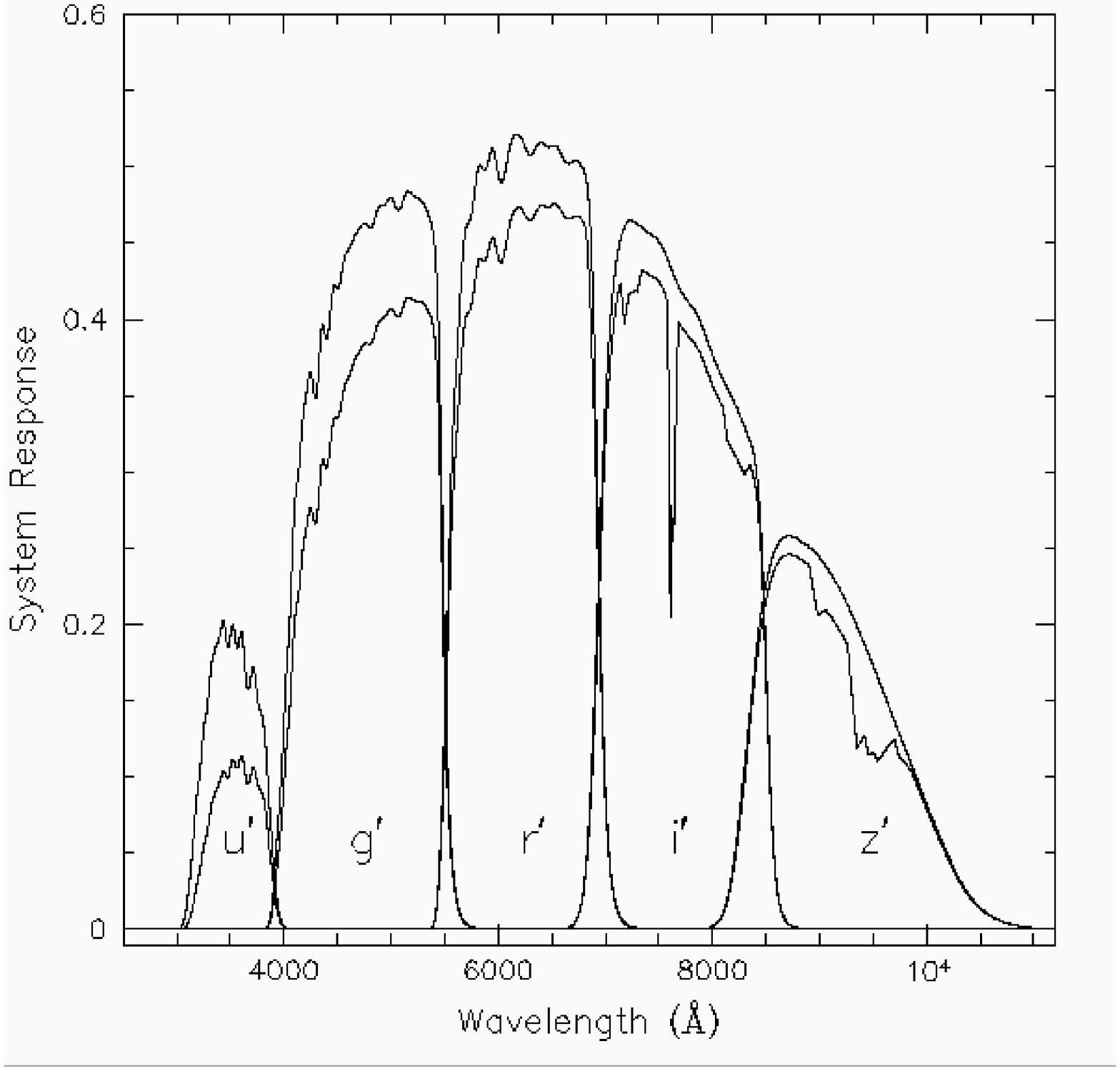}
\caption{SDSS photometric system response as a function of wavelength in
Angstroms. 
The upper curve is without atmospheric extinction, the lower curve
includes the effects of atmospheric extinction when observing at a
typical altitude of 56\dg.}
\label{fig:filters}
\end{figure}

The SDSS photometric system $u'g'r'i'z'$ \cite{fuku96} 
(Figure~\ref{fig:filters}) has been 
specifically designed for this survey and covers the near-UV to near-IR range
($\sim 3000$--$10,000$\AA) in five essentially non-overlapping passbands.
The $u$ filter response peaks in the near ultra-violet at 3500\AA, 
$g$ is a blue-green band centred at 4800\AA, $r$ is a red band centred at
6250\AA, $i$ is a far-red filter centred at 7700\AA\ and $z$ is a near-infrared
passband centred at 9100\AA.
The standard stars that define this system have been presented by Smith et al.\
\cite{smith2002}.
The photometric data are not yet
finally calibrated, so the current magnitudes are indicated with asterisks,
$u^*g^*r^*i^*z^*$, to denote their preliminary nature.
The SDSS filters themselves are referred to simply as $ugriz$, 
without primes or asterisks.

In order to provide photometric calibration while the imaging camera is
scanning, a second, dedicated {\em Photometric Telescope} operates
concurrently, observing photometric standard stars and creating photometrically
calibrated ``secondary patches'' which lie within the main telescope's
scan.  These calibration patches are then used to transfer the
primary photometric calibration to objects detected with the 2.5m telescope
and imaging camera.
The photometric quality of the data is monitored by a software ``robot''
that automatically rejects data observed during cloudy periods \cite{hfsg2001}.

The other 24 CCDs in two additional dewars are also SITe chips of width
2048  24$\mu$ pixels,
but they have only 400 rows in the scanning direction.  These dewars are
oriented perpendicular to the photometric dewars, with one at the
top and one at the bottom of the imaging array.
Two of these CCDs (one in each dewar)
are used to determine changes in focus.  The remaining 22 CCDs
reach brighter magnitudes before saturating and are
used to tie our observations to an astrometric reference frame
defined by bright stars which saturate our imaging detectors \cite{pier2002}.

\begin{figure}
\includegraphics[angle=-90,width=\textwidth]{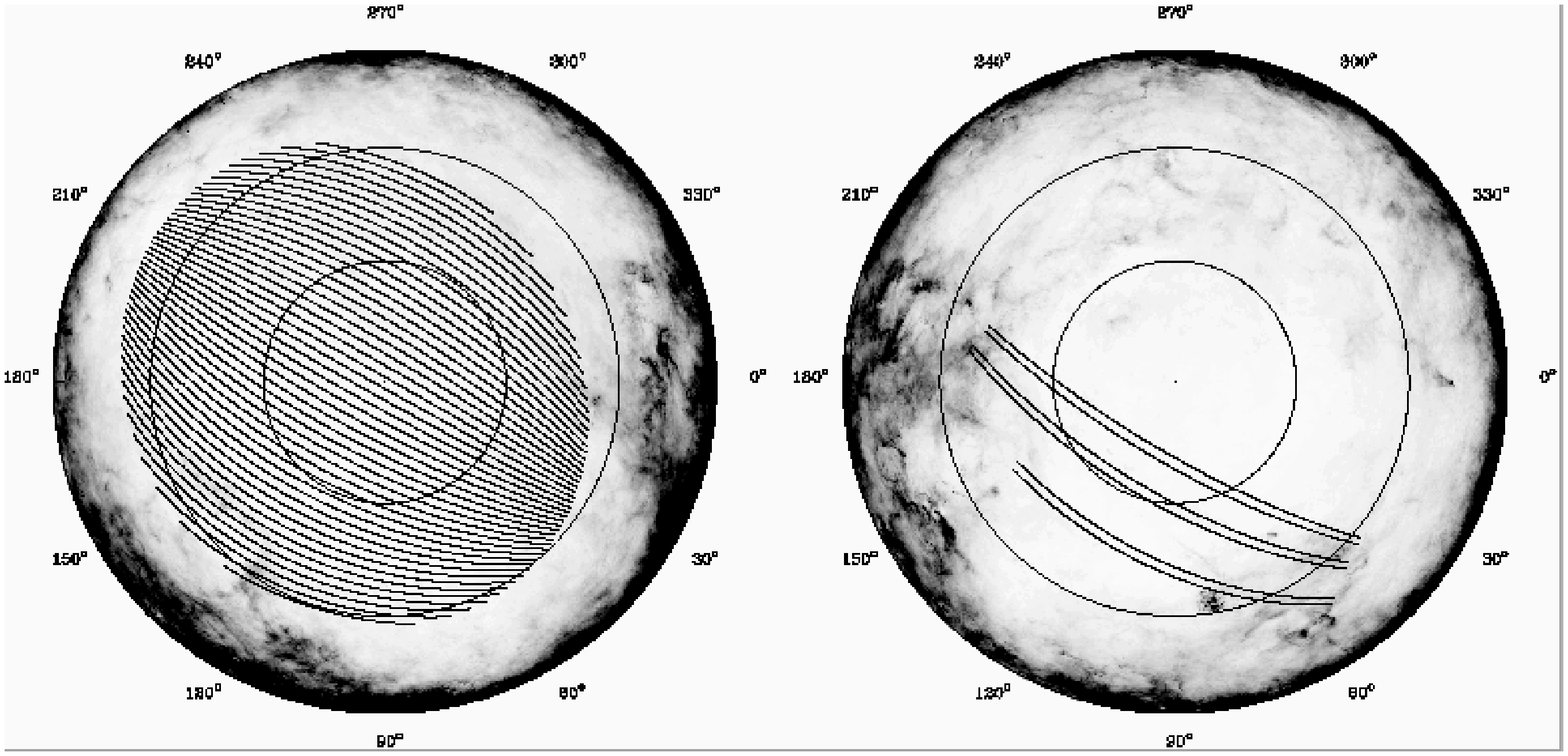}
\caption{The location of the survey imaging stripes plotted in a polar
projection for the North (left) and South (right) Galactic hemispheres.
The grey scale indicates the amount of reddening due to dust in our own
Galaxy \cite{sfd98}, where white corresponds to no reddening and black
to one magnitude of reddening in $g-r$ colour.
The circles are lines of constant Galactic latitude ($|b| = 30\dg, 60\dg$),
and Galactic longitude is marked around the edge of each hemisphere.}
\label{fig:stripes}
\end{figure}

The location of the survey imaging area is shown in Figure~\ref{fig:stripes}.
The northern survey area is centred near the North
Galactic Pole and it lies within a nearly elliptical shape $130\dg$ E-W
by $110\dg$ N-S chosen to minimize Galactic foreground extinction.
All scans are conducted along great circles in order to
minimize the transit time differences across the camera array.  
There are 45 great circles (``stripes'') in the northern survey region 
separated by 2.5\dg.
We observe three non-contiguous stripes in the Southern Galactic Hemisphere,
at declinations of 0, $+15\dg$ and $-10\dg$, during parts of the autumn
season when the northern sky is unobservable.
Each stripe is scanned
twice, with an offset perpendicular to the scan direction in order
to interlace the photometric columns.  
A completed stripe slightly exceeds $2.5\dg$ in width and thus there 
is a small amount of overlap to allow for telescope mis-tracking and to 
provide multiple observations of some fraction of the sky for quality 
assurance purposes.
The total stripe
length for the 45 northern stripes will require a minimum of 650 hours of
pristine photometric and seeing conditions to scan at a sidereal rate.
Based upon our current experience of observing conditions at APO, it
seems likely that we will only complete about 75\% of this imaging
after 5 years of survey operations.

\subsection{Spectroscopic Survey}

The goals of the spectroscopic survey are to observe spectra for 
$10^6$ galaxies, $10^5$ quasars and $10^5$ stars.
In order to obtain the spectra of over $10^6$ objects in a survey
covering 10$^4$ square degrees, we must obtain spectra of about 100
objects per square degree.  Although some overlap of fields is
inevitable, we would like to keep this overlap to a minimum for reasons
of efficiency and cost.  Hence, we need
to obtain several hundred spectra per 3\dg\ diameter spectroscopic field.

To accommodate this requirement, two identical multi-fibre spectrographs
have been built which are each fed by 320 fibres.  The spectrographs
cover the wavelength range 3900--9100\AA\ at a resolution of 
$\lambda/\Delta\lambda \sim$ 1800,
or 167 km s$^{-1}$.  Each spectrograph has two cameras, one optimized for
the red and the other for the blue.  Each camera has as its detector
a $2048 \times 2048$ CCD with 24$\mu$ pixels.

The 180$\mu$ fibres, which each subtend $3''$ on the sky,  
are located in the focal plane by 
plugging them by hand into aluminium plates which are precisely drilled for
each field based upon the astrometric solution obtained from the imaging
data. To avoid mechanical interference, individual fibres can be placed
no closer than $55''$ to one another.
The plates and fibres are held in the focal plane, and coupled with
the spectrographs, by one of 9 identical rigid assemblies called cartridges.
Since all of the cartridges can be pre-plugged during the day, 5,760 spectra
can be obtained during a long night without re-plugging.  A mapping procedure
is invoked after plugging each cartridge that automatically tags each fibre
to the appropriate object on the sky.

A surface density of 100 galaxies per square degree corresponds roughly
to an $r$-band magnitude limit $r \approx 18$.
To obtain redshifts for galaxies of this magnitude
requires exposure times of about 45 minutes,
which we split into three, 15 minute exposures to aid in rejection of
cosmic rays.
Cosmic ray events occur at essentially random locations on our detectors,
and so are very unlikely to appear at the same place in all three exposures.
Each field takes about one hour, including calibration
(flat field and comparison lamp) exposures and allowing for telescope
pointing and the exchange of fibre cartridges.
Spectroscopic observations are carried out whenever
observing conditions are not adequate for imaging, ie. when seeing exceeds
1.5 arcsec or when skies are non-photometric.

\subsection{Data Processing}

All of the raw data from the photometric CCDs are archived.
The frames are first read to disk, then written to DLT tape.  
Over 16 Gb per hour are generated from the
photometric chips.  When observing in spectroscopic mode, 
the amount of data generated seems trivial in
comparison (about 6 exposures per hour for each of two cameras for each
of two spectrographs, or 24 8Mb frames per hour).

All data tapes are shipped by overnight express courier to Fermi National
Accelerator Laboratory, near Chicago, where the data reduction
pipelines are run.
The goal is to turn the imaging data around within a few
days, so that one dark run's worth of imaging data will be processed before
the next dark run begins, allowing objects to be targeted for spectroscopy.
The data flow serially through several pipelines to identify, measure and
extract astronomical images and to
apply photometric and astrometric calibrations.
Once a significant area of sky has been imaged, a {\em target selection}
procedure is then run in order to select objects for followup spectroscopy.
Next, an {\em adaptive tiling} algorithm 
assigns targets to spectroscopic plates and chooses plate centres in
order to maximize observing efficiency \cite{blmyzl2001}.
Since galaxies are clustered on the sky, the target density varies from
place to place.
In regions of high target density, the adaptive tiling algorithm allows
the plates to move slightly closer together so that all targets can be observed.
The plates are then manufactured and shipped to the site.

Spectroscopic reduction is also automated.
We are able to obtain correct redshifts for 99\% of targeted objects, 
without human intervention.
The pipelines are integrated into a specially-written environment
known as Dervish,
and the reduced data are stored in an object-oriented database.

\subsection{Spectroscopic Samples}

There are several distinct spectroscopic samples observed by the survey.
In a survey of this magnitude, it is important that the selection 
criteria for each sample remain fixed throughout the duration of the survey.
Therefore, we spent a whole year obtaining
test data with the survey instruments and refining the spectroscopic selection
criteria in light of our test data.
Now that the survey proper is underway, these criteria have been 
``frozen in'' for the duration of the survey.

The {\bf main galaxy sample} \cite{strauss2002}
consists of $\sim 900,000$ galaxies selected
by $r$ band magnitude, $r^* < 17.77$.
This magnitude limit was chosen as test year data demonstrated that it
corresponds closely to the desired target density of 90 objects per
square degree.
Since galaxies are fuzzy, extended sources, there is no easy way to measure
their total magnitude.
Most previous surveys have measured the light within an isophote of constant
surface brightness, but these isophotal magnitudes will systematically
underestimate the flux of galaxies of low intrinsic surface brightness 
and at high
redshift $z$, since observed surface brightness scales as $(1+z)^4$.
Simulations have shown that the Petrosian magnitude \cite{pet76},
which is based on an aperture defined by the ratio of light within an annulus 
to total light inside that radius, provides probably the least biased
and most stable estimate of total magnitude.
We therefore select galaxies according to their Petrosian magnitude.
We also apply a surface-brightness limit, $\mu_{r^*} < 24.5$ mag arcsec$^{-2}$,
so that we do not
waste fibres on galaxies of too low surface brightness to give a reasonable
spectrum.
This surface brightness cut eliminates just 0.1\% of galaxies that would
otherwise be selected for observation.
Galaxies in this sample have a median redshift 
$\langle z \rangle \approx 0.104$.

We will observe an additional $\sim 100,000$ {\bf luminous red galaxies} 
\cite{eisen2001}.
Given photometry in the five survey bands, redshifts can be estimated
for the reddest galaxies to $\Delta z \approx 0.02$ or better,
and so one can also predict their intrinsic luminosities quite accurately.
The luminous red galaxies, many of which will be so-called central dominant
(cD) galaxies in cluster 
cores, provide a valuable supplement to the main
galaxy sample since 1) they have distinctive spectral features,
allowing a redshift to be measured for objects to a flux limit
around 1.5 magnitudes fainter than the main sample,
and 2) they form a volume-limited sample, ie. a sample of uniform density,
out to redshift $z = 0.38$.
This sample will thus be extremely powerful for studying
clustering on the largest scales and for investigating galaxy evolution.

{\bf Quasar} candidates \cite{richards2002} are selected from cuts
in multi-colour space and by identifying sources 
from the FIRST radio catalogue \cite{bwh95},
with the aim of observing $\sim 100,000$ quasars.
This sample will be orders of magnitude larger than any existing quasar
catalogue, and will be invaluable for quasar luminosity function, evolution
and clustering studies
as well as providing sources for followup absorption-line observations.

In addition to the above three classes of spectroscopic sources, which are
designed to provide {\em statistically complete} samples, we are also 
obtaining spectra
for many thousands of {\bf stars} and for various {\bf serendipitous}
objects.
The latter class includes objects of unusual colour or morphology
which do not fit into the earlier classes, plus unusual objects found
by other surveys and in other wavebands.

\subsection{Survey Status}

First light with the imaging camera was obtained on 9 May 1998 and 
the first extra-galactic spectra were obtained in June 1999.
The survey officially began on 1 April 2000, and observing is scheduled
to end on 30 June 2005.
At the time of writing (June 2002), we have imaged \area\ square degrees
(\pcarea\% of the total survey area) and obtained spectra for 
\plate\ plug-plates,
yielding spectra for \galspec\ galaxies, \qsospec\ quasars and \starspec\ stars,
including some repeated observations.
The spectrographs are performing extremely efficiently, with an overall
throughput, including telescope optics but excluding the atmosphere,
of 20\% in the blue (3900--6000 \AA) and 25\% in the red
(6000--9100 \AA).
Automated spectral reduction pipelines classify these spectra and measure
redshifts.
Conservatively, we inspect the spectra of roughly 8\% of sources, whenever
there is any doubt about the reliability of the automated redshift 
measurement.
In seven-eighths of these cases, the automated redshift measurement is in 
fact confirmed to be correct.
The remaining eighth of these spectra (1\% overall) have their redshifts
manually corrected.
Based on manual inspection of all $\approx 23,000$ spectra from 39 plugplates,
this procedure correctly measures redshifts
for 99.7\% of galaxies, 98.0\% of quasars and 99.6\% of stars.

\section{The Early Data Release}
\label{sec:edr}

The first public release of SDSS data (hereafter EDR) took place on 
5 June 2001, and consists
of images covering 460 square degrees of sky, photometric parameters for
10 million objects and spectra for 54,000 objects.
The main access point to the data is through the website 
{\tt http://www.sdss.org/} and the EDR is described in
\cite{stough2002}.

There are presently three ways to access the data, the choice of which depends
on the nature of the data required and the experience of the user.

The {\bf SkyServer} ({\tt http://skyserver.fnal.gov}) provides a graphical
user interface to images of the sky and also enables one to download spectra
of specified objects.
It was primarily intended as an interface for the general public and for
educational purposes, but new features are being added, making it also useful
for professional astronomers.
Public interest in the SDSS is illustrated by the fact that this web site 
has been receiving around half a million hits per month.

The {\bf MAST} interface ({\tt http://archive.stsci.edu/sdss/})
allows simple web-based searches around specified
objects or positions on the sky.
It is a useful way of retrieving SDSS observations for moderate
numbers of objects in a small region of the sky.

For accessing SDSS data on large numbers of objects, and over larger areas,
the SDSS query tool {\bf sdssQT} is recommended.
This tool allows one to query the EDR database on any measured parameters
and to specify which parameters, such as position and magnitude, are to be
returned.
Documentation on the query tool is available from
{\tt http://archive.stsci.edu/sdss/software/}.

\begin{figure}
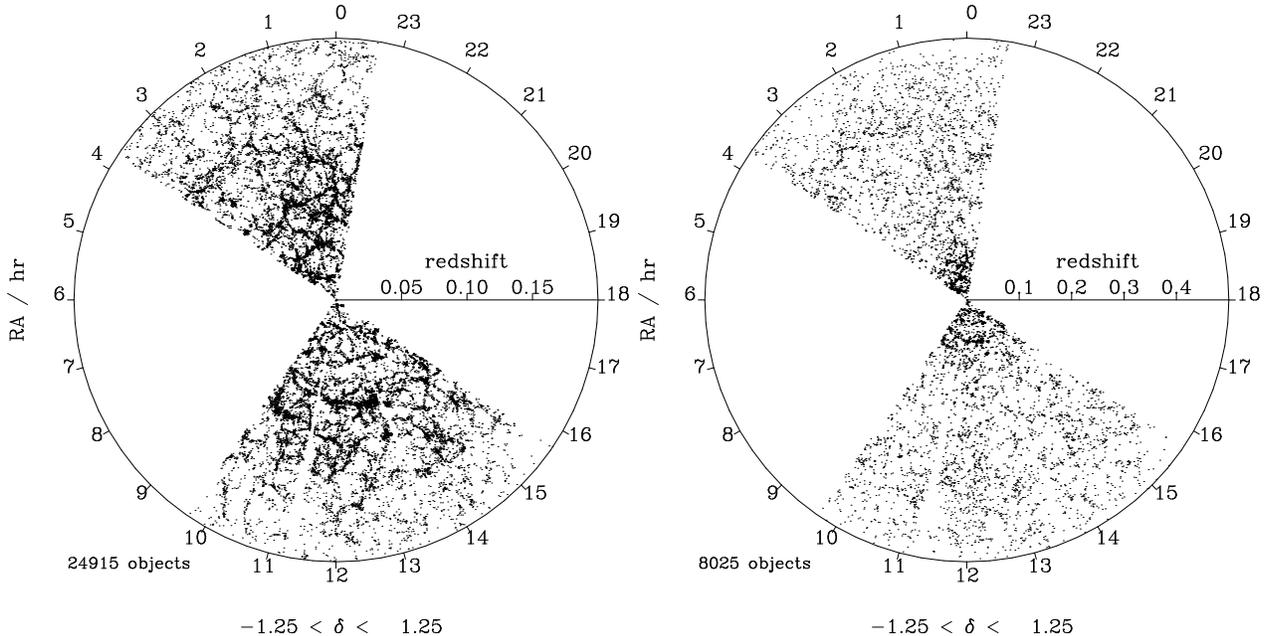

\includegraphics[width=0.5\textwidth]{zMain02}
\hfil
\includegraphics[width=0.5\textwidth]{zLRG05}
\caption{Distribution of EDR galaxies in right ascension (RA) and redshift 
around the equator (declination $|\delta| < 1.25\deg$).
The left plot shows 24,915 galaxies from
the main flux-limited galaxy sample within
a redshift $z=0.2$.
The right plot shows 8025 galaxies from the luminous red galaxy sample 
to $z = 0.5$.}
\label{fig:wedge}
\end{figure}

The distribution of equatorial galaxies in
the EDR is shown in right ascension (RA) versus redshift wedge plots in 
Figure \ref{fig:wedge}.
The main galaxy sample is flux-limited ($r^* < 17.6$) and has a median redshift
$\bar{z} \approx 0.11$.
The clustering of galaxies is clearly visible in this plot:
the galaxies appear to lie within filamentary structures enclosing regions
of substantially lower density.
The drop in galaxy density with redshift (distance from the centre of the plot)
is entirely due to the fact that this sample is limited by apparent flux:
only the most luminous galaxies, which are rare, can be seen beyond a redshift
$z \gtrsim 0.15$.

By contrast, the luminous red galaxy (LRG) sample is designed to be 
volume-limited, ie. to be of uniform density, out to redshift $z = 0.38$.
This sample also includes additional galaxies to $z \sim 0.5$, although these
high redshift galaxies do not form a complete subsample.
At $z < 0.15$, the simple linear colour cut used allows less luminous 
galaxies to enter the sample, hence the increase in galaxy density at 
these low redshifts.

The EDR includes some engineering data of sub-survey standard, and
will be superseded by the first official data release, DR1, in January 2003.
This release will include spectra of more than 200,000 objects over 
2800 square degrees of sky.
Subsequent data releases will follow at roughly yearly intervals.

\section{Early Science Results}
\label{sec:science}

Although the primary science driver behind the Sloan Digital Sky Survey
is characterization of the large scale structure of the Universe,
the survey has already had a significant impact on several branches of
astrophysics, from the investigation of asteroids in our own Solar System
to the discovery of the most distant known objects in the Universe.
Here I very briefly highlight some interesting results which have come 
out of the commissioning phase of the survey.
For further details, please see the original articles as referenced.

\subsection{Asteroids}

Asteroids are easily detected in SDSS imaging data since they are
fast-moving and very nearby (within the Solar System), leading to a significant
motion relative to the background stars during the 55 s integration time.
They thus appear on SDSS images as trails, the length and orientation of which
enable the asteroid's orbit to be determined.
Around 13,000 asteroids have been detected in 500 deg$^2$ of SDSS commissioning
data \cite{ivezic2001}.
These observations have enabled an accurate determination of the size
distribution of asteroids to $r^* < 21.5$ over the range 0.4--40 km.
The total number of predicted asteroids with $r^* < 21.5$ is about a factor
of ten smaller than that predicted by an extrapolation from previous
observations of brighter asteroids ($r^* \lesssim 18$), and the number of
``killer'' asteroids with diameter $D > 1$ km is a factor of about three 
smaller than previously thought, with a new estimate of roughly one impact
per 500,000 years.
By completion, we estimate that the SDSS will have observed roughly 100,000
asteroids in five colours, enabling their approximate chemical composition
to be determined.
There is already clear evidence for chemical segregation in the belt of 
asteroids between Mars and Jupiter.
Asteroids in the inner part of the belt are composed mostly of
rocky silicates, whereas the outer belt asteroids are primarily carbonaceous.
These observations have important implications for the formation history 
of the Solar System.

\subsection{Brown Dwarfs and Methane Dwarfs}

Moving slightly further afield than the Solar System, the SDSS has also been
very successful at finding brown dwarfs in the vicinity of the Sun.
Brown dwarfs are sub-stellar objects which are too small to sustain
thermonuclear reactions in their cores, and are thus not true stars,
but are larger than planets
(they are thought to be 10--70 times the mass of Jupiter).
%These stars are at the cool end of the stellar sequence, of type M, L and T.
%\footnote{Traditionally, the stellar sequence
%has been represented by the letters OBAFGKM, from hot to cool stars.
%Recently, the discovery of objects cooler and fainter than class M has resulted
%in the addition of two new classes, L and T.}
They are thus cool, below 2,500 K, and hence very red, enabling them to 
be easily distinguished from true stars by their colours in the SDSS filters.
To date \cite{hawley2002}, SDSS has discovered around fifty new brown dwarfs,
including four T-dwarfs, also known as methane dwarfs.
The methane dwarfs are so cool, below 1,300 K,  that their spectra are 
dominated by the presence of molecules such as water vapour and methane.
While a methane dwarf, Gliese 229B, had already been discovered in orbit
around a brighter star, the SDSS was the first survey to discover 
free-floating methane dwarfs.
Using the discovery technique pioneered by SDSS, the Two Micron All Sky Survey
(2MASS) has since found more than a dozen methane dwarfs \cite{burg2001}.
Although very low in mass, these elusive objects may be very common,
and thus may provide a significant fraction of the ``dark matter'' that is
known to exist in the Milky Way.
To understand their contribution to the total mass of the Milky Way requires
determining both their abundance and their mass.
The SDSS will be important in addressing both these questions, due to its
capability of obtaining precision five-colour photometry over a very large 
area of sky.

\subsection{Star Structures in the Galaxy}

The SDSS has discovered an unexpectedly large number of blue stars
within 20 degrees of the Galactic plane.
It is thought that these stars could be part of a disrupted dwarf galaxy,
or a disk-like distribution of stars that is puffier than accepted models 
of the stellar disk of the Galaxy, and flatter than the spherical
distribution in the halo \cite{newb2002}.
These observations suggest that the model for our Galaxy needs to be 
reconsidered.
One possible explanation is that these star structures came from
tidal disruption of nearby dwarf galaxies such as Sagittarius,
since the ages and metallicities of the stars are consistent with the
stellar populations in Sagittarius.

\subsection{Galaxy luminosity function}
\label{sec:lf}

The distance to a galaxy can be obtained from its spectroscopic redshift
using Hubble's law, which says that the recession velocity of a galaxy is 
linearly proportional to its distance.
Knowing the distance to a galaxy, its intrinsic luminosity may be
determined from its apparent magnitude using the inverse-square law.
By calculating the maximum distance to which a galaxy of given luminosity
may be seen, one can find the {\em galaxy luminosity function}, $\phi(L)$,
the number density of galaxies as a function of intrinsic luminosity.
It has long been known that the galaxy LF is
well fit by the Schechter function \cite{schec76},
$$
\phi(L)\; dL = \phi^*\bigg ( {L \over L^*} \bigg )^{\alpha}\; {\rm exp}\bigg
( -{L \over L^*} \bigg )\; d\bigg( {L \over L^*} \bigg),
\label{eqn:sch}
$$
in which 
the number density of faint galaxies is described by a power law with
index $\alpha \approx -1.2$.
For galaxies brighter than a characteristic luminosity $L^*$, the number
density drops exponentially.

We have determined the luminosity function from a sample of 11,275 galaxies
in the SDSS commissioning data \cite{blan2001}, 
and find that the LF is extremely well fit
by a Schechter function over a range of 8 magnitudes.
The use of Petrosian magnitudes by the SDSS means that a larger fraction
of a galaxy's light is being measured than is the case with most previous 
surveys.
The integrated {\em luminosity density} of the Universe determined from
SDSS measurements is thus a factor 1.5--2 times larger than previously thought.
This improves the agreement with most models of galaxy formation, 
which predict a higher density of stellar matter than previous observational
estimates.

\subsection{Galaxy clustering}

The clustering of galaxies provides a powerful probe of the initial density
perturbations in the early Universe, the power spectrum of which is
determined by cosmological parameters such as the matter density and Hubble
parameters, $\Omega_m$ and $h$ respectively.
On large scales, $r \gtrsim 20$ Mpc, the clustering of galaxies is related
to the primordial density fluctuations by linear perturbation theory,
but on smaller scales, non-linear effects can change the shape of the observed
clustering pattern.

\begin{figure}
\includegraphics[width=0.5\textwidth]{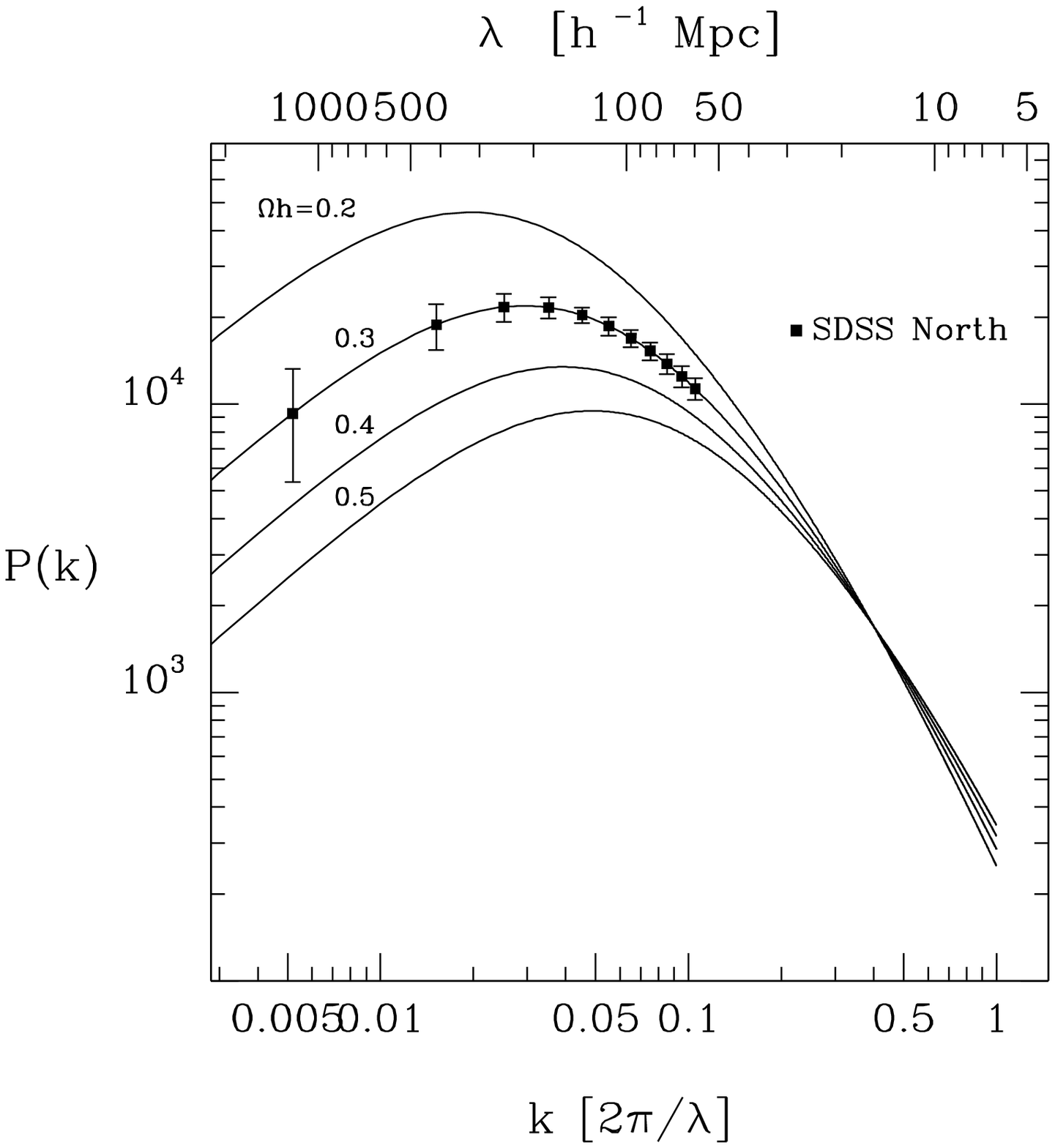}
\hfil
\includegraphics[width=0.5\textwidth]{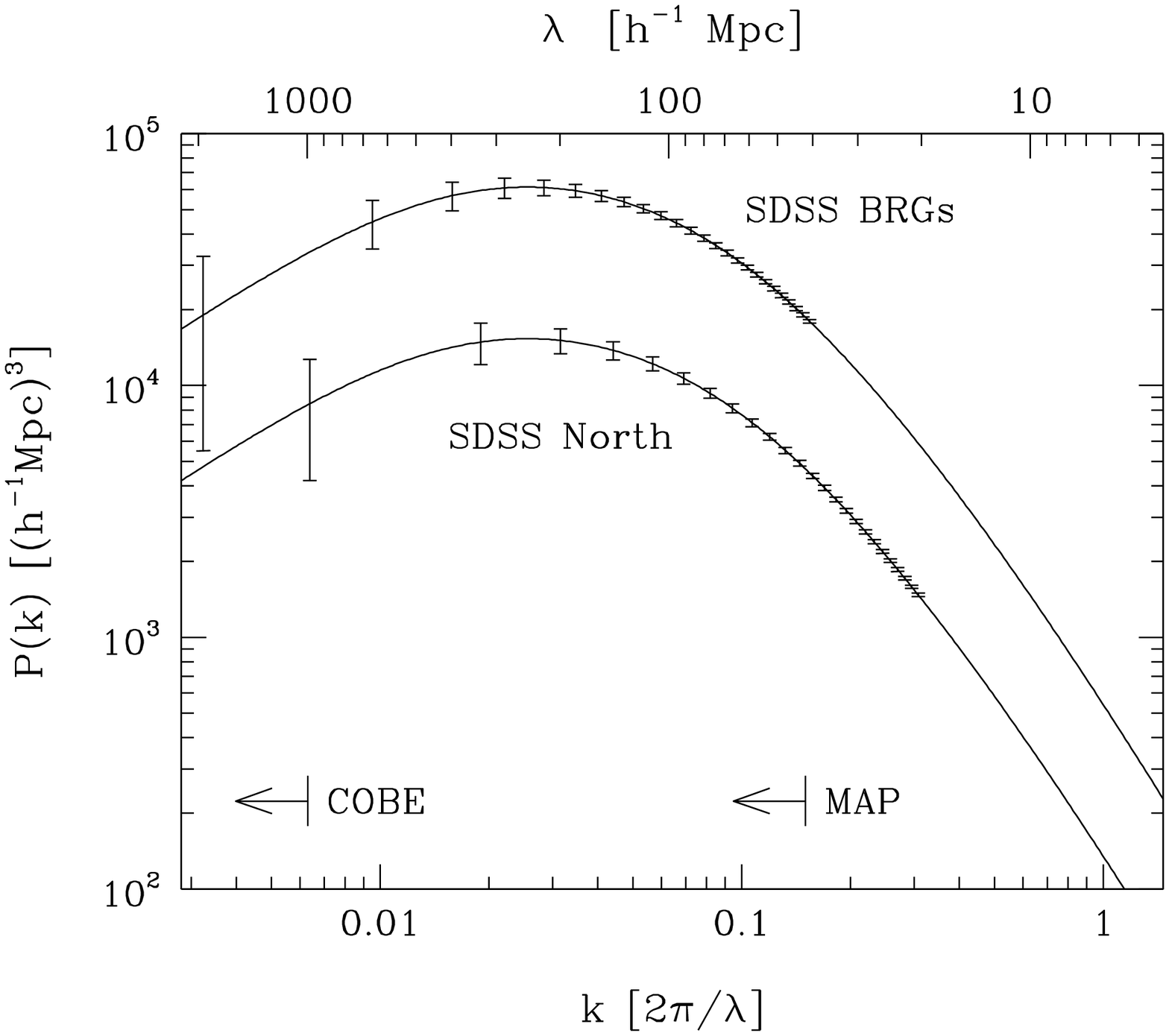}
\caption[]{Left: (a) Expected $1\sigma$ uncertainty in the galaxy power spectrum
$P(k)$ we would measure from a volume-limited sample from the completed
SDSS northern survey,
along with predictions of $P(k)$ from four variants of the
low-density CDM model.
Note that the models have been arbitrarily normalized to agree
on small scales ($k = 0.4$); in practice the COBE observations
of CMB fluctuations fix the amplitude of $P(k)$ on very large scales.
Right: (b) Power spectrum expected from the luminous red galaxy sample
(BRGs), assuming that these galaxies are four times as strongly
clustered as the main sample galaxies.}
\label{fig:P_k}
\end{figure}

The clustering of galaxies is most frequently measured with
the two-point correlation function $\xi(r)$, which gives the excess probability
above random of finding two galaxies at separation $r$, or its Fourier inverse,
the power spectrum $P(k)$.
The huge volume of the completed SDSS redshift survey will enable
estimates of the galaxy power spectrum to $\sim 1000 \hMpc$ scales.
Figure~\ref{fig:P_k}a shows the power spectrum $P(k)$ we would expect
to measure from a volume-limited sample of galaxies from
the SDSS northern redshift survey, assuming Gaussian fluctuations
and a $\Omega_m h = 0.3$ cold dark matter (CDM) model.
The error bars include cosmic variance and shot noise, but not systematic
errors, due, for example, to Galactic obscuration.
Provided such errors can be corrected for, we can easily distinguish 
between $\Omega_m h = 0.2$, as indicated by a recent study of the masses
of SDSS galaxy clusters \cite{bahcall2002}, and $\Omega_m h = 0.3$,
favoured by previous studies, as Figure~\ref{fig:P_k}a demonstrates.
The noted systematic effects of Galactic obscuration will be minimized by
SDSS observations of distant F stars, far from the Galactic centre.
These stars, three of which are observed on each spectroscopic plate,
have well-understood intrinsic colours, and so can be used as 
reliable reddening indicators.
Adding the luminous red galaxy sample
(Fig.~\ref{fig:P_k}b), will
further decrease measurement errors on the largest scales, and so we also
expect to be able to easily distinguish between low-density CDM and 
mixed dark matter (MDM) models,
and models with differing shapes of the primordial
fluctuation spectrum.

Given the relatively small sky coverage of our existing data, we are not
yet able to place reliable constraints on cosmological parameters.
We can however use observations of galaxy clustering to test the
quality of our photometry and star-galaxy separation, and to investigate
the {\em relative} clustering strength of different types of galaxies
on small scales.

\subsubsection{Angular clustering}

A series of papers \cite{conn2001,dod2001,scran2001,szalay2001,teg2002}
have studied the angular clustering of galaxies in SDSS commissioning data.
These papers are based on a single survey stripe (runs 752/756 observed in
March 1999) measuring $2.5 \times 90$ degrees and containing some 3 million
galaxies to $r^* = 22$.
Star-galaxy separation is performed using a Bayesian likelihood 
and approximately 30\% of the area is masked out due to poor seeing
\cite{scran2001}.
The angular correlation function, $w(\theta)$, which describes the excess 
probability over random of finding two galaxies at angular separation 
$\theta$, is consistent with that 
measured from the APM Galaxy Survey \cite{mesl90} when scaled to the
same depth \cite{conn2001}.

\begin{figure}
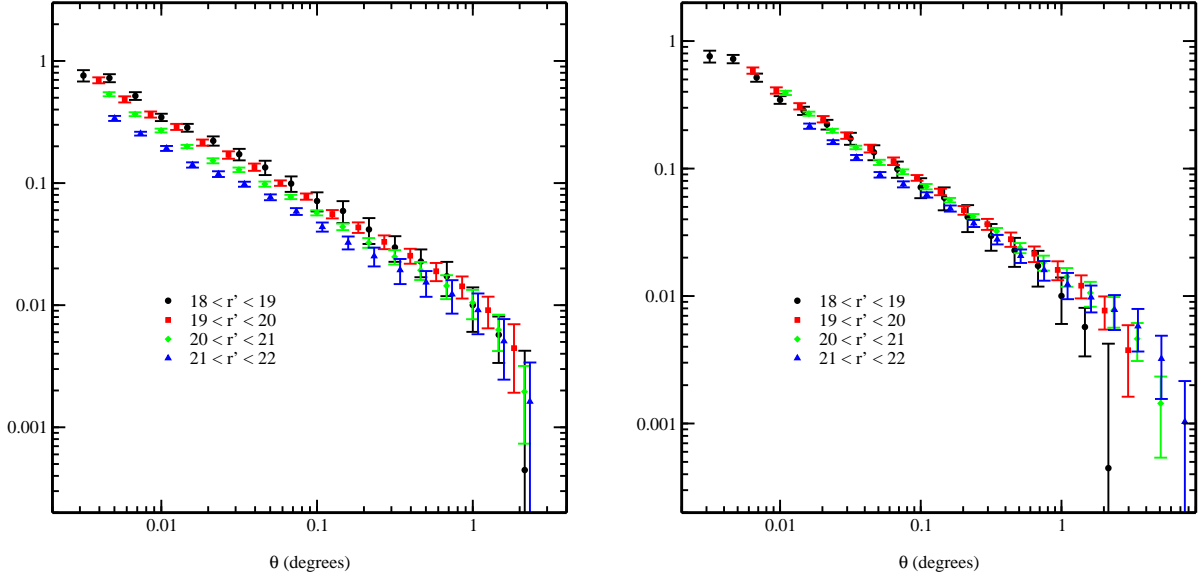

\includegraphics[width=0.45\textwidth]{wslice}
\hfil
\includegraphics[width=0.45\textwidth]{wscale}
\caption{Scaling of the angular correlation function $w(\theta)$
measured in four
magnitude slices according to Limber's equation (see text) and assuming a
selection function assumed based on the CNOC2 galaxy redshift survey 
\cite{lin99}.
The left plot assumes a $\Omega_m = 1$, $\Omega_\Lambda = 0$ cosmology,
the right plot a $\Omega_m = 0.3$, $\Omega_\Lambda = 0.7$ cosmology.
Assuming the latter cosmology improves the scaling of the faintest
($21 < r' < 22$) slice, since the volume per unit redshift is larger
at high redshifts in the $\Lambda$-dominated cosmology.
From~\cite{scran2001}.}
\label{fig:limber}
\end{figure}

An important test of the star-galaxy separation and of the photometric 
calibration is to check that $w(\theta)$ scales as expected with apparent
magnitude.
(We expect $w(\theta)$ to shift to smaller angular scales and a
lower amplitude as we look at more distant, and hence apparently fainter, 
galaxies.  This scaling is quantified by Limber's equation \cite{limber53}.)
Figure \ref{fig:limber} shows that the scaling of $w(\theta)$ is well-described
by Limber's equation, particularly when a vacuum-dominated ($\Omega_m = 0.3$, 
$\Omega_\Lambda = 0.7$) cosmology is assumed.
Further tests for possible sources of systematic errors in the SDSS data
are described in detail in \cite{scran2001} and the angular clustering 
results are summarized in \cite{conn2001}.
% Other papers describe estimates of the angular power spectrum \cite{teg2002},
% inversion from $w(\theta)$ to the 3-dimensional power spectrum $P(k)$ 
% \cite{dod2001} 
% and direct estimation of power spectrum parameters \cite{szalay2001}.

\subsubsection{Spatial clustering}

\begin{figure}
\hspace{-10mm}
\includegraphics[width=0.6\textwidth]{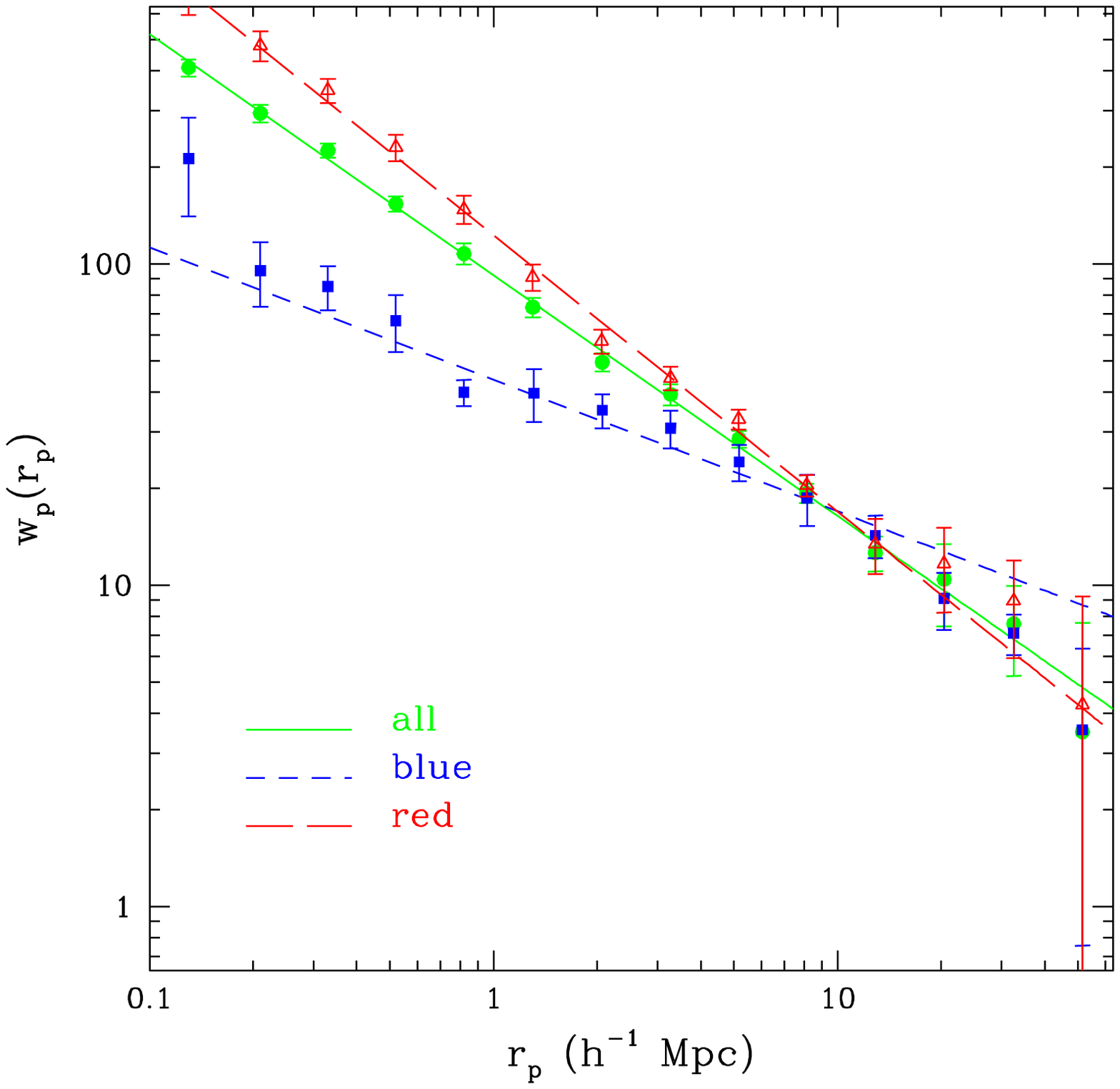}
\hspace{-15mm}
\includegraphics[width=0.6\textwidth]{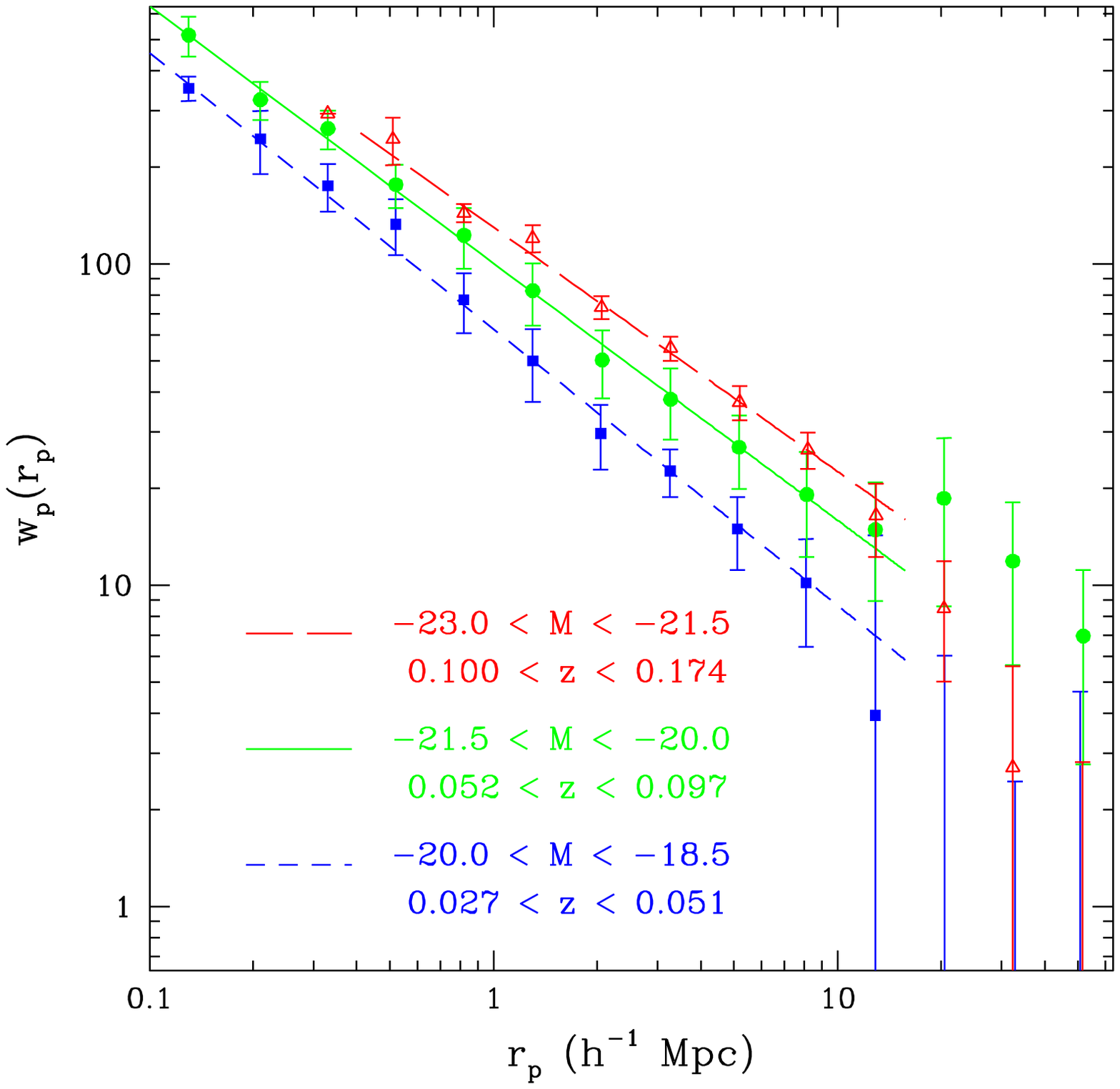}
\caption{Projected correlation functions $w_p(r_p)$ against projected
separation $r_p$ for redshift survey
galaxies subdivided by colour (left plot) and luminosity (right plot).
Note that the slope of $w_p(r_p)$ increases from blue to red colour,
but remains approximately constant with luminosity.
From~\cite{zehavi2002}.}
\label{fig:xi}
\end{figure}

A preliminary estimate of the spatial clustering of galaxies has been made
using redshift information \cite{zehavi2002}.
This sample consists of 29,300 galaxies with $r^* < 17.6$ and within $\pm 1.5$
magnitudes of the characteristic magnitude $M_r^*$ (corresponding to the
characteristic luminosity $L^*$ in the Schechter function fit to the
luminosity function, see \S\ref{sec:lf}), distributed 
non-contiguously over 690 square degrees.

When using redshifts to infer distances, one relies on the Hubble relation,
ie. that distance is proportional to recession velocity.
In fact, galaxies have peculiar velocities relative to the Hubble expansion,
leading to an error in estimated distances.
It is important to take these distance errors, or redshift-space distortions,
into account when measuring galaxy clustering.
One way of doing this is to estimate galaxy clustering as a function 
$\xi_2(r_p, \pi)$ of two components of the separation vector: 
the line of sight separation $\pi$, which is affected by peculiar velocities,
and the sky-projected separation $r_p$, which is not.
In the absence of redshift-space distortions, 
the contours of $\xi_2(r_p, \pi)$ 
would be symmetric about the origin, but small-scale peculiar velocities
cause an elongation of the contours along the line of sight direction $\pi$,
the so-called ``finger of God'' effect.
One can estimate a projected correlation function $w_p(r_p)$ that is
unaffected by redshift-space distortions by integrating $\xi_2(r_p, \pi)$
along the line of sight $\pi$,
$$
w_p(r_p) = 2 \int_0^\infty d\pi \xi_2(r_p, \pi) 
= 2 \int_0^\infty dy \xi(\sqrt{r_p^2 + y^2}),
$$
where the second integral relates $w_p(r_p)$ to the spatial correlation
function $\xi(r)$.

We find that $\xi(r)$ is well-fit over the range $0.1 < r < 30 \hMpc$
by a power law $\xi(r) = (r/r_0)^{-\gamma}$ with parameters given in
Table~\ref{tab:corr}.
The correlation length $r_0 = 6.14 \pm 0.18\hMpc$ is larger than
$r_0 = 5.1 \pm 0.2 \hMpc$ found by an earlier study \cite{lmep95}, 
presumably because dwarf
galaxies with $M > M^* + 1.5$ have been excluded from the SDSS analysis.
The index $\gamma = 1.75 \pm 0.03$ is in very good agreement with the
earlier result ($\gamma = 1.71 \pm 0.05$).

This analysis also allows one to explore the dynamics of galaxies.
We find that for close pairs of galaxies, at a projected separation 
$r_p < 5 \hMpc$, the rms relative velocity of galaxies 
$\sigma \approx$ 600 km/s.

\begin{table}
\begin{center}
\caption{Power-law parameters for the real-space correlation function
$\xi(r) = (r/r_0)^{-\gamma}$.  Units for the correlation length $r_0$
are $\hMpc$.  From~\cite{zehavi2002}.}
\label{tab:corr}
\vspace{5mm}
\begin{math}
\begin{array}{lll}
\hline
\hline
\mbox{Sample} & \multicolumn{1}{c}{r_0} & \multicolumn{1}{c}{\gamma}\\
\hline
\mbox{All} & 6.14 \pm 0.18 & 1.75 \pm 0.03\\
\mbox{Red} & 6.78 \pm 0.23 & 1.86 \pm 0.03\\
\mbox{Blue} & 4.02 \pm 0.25 & 1.41 \pm 0.04\\
M^* - 1.5   & 7.42 \pm 0.33 & 1.76 \pm 0.04\\
M^*         & 6.28 \pm 0.77 & 1.80 \pm 0.09\\
M^* + 1.5   & 4.72 \pm 0.44 & 1.86 \pm 0.06\\
\hline
\hline
\end{array}
\end{math}
\end{center}
\vspace{-5mm}
\end{table}

Figure~\ref{fig:xi} shows the clustering properties for two subsamples of the
galaxy population selected by restframe $u-r$ colour at $(u^*-r^*)_0 = 1.8$,
corresponding roughly to bulge (red) and disk (blue) dominated galaxies.
The red galaxies exhibit a steeper power-law slope and longer correlation
length than the blue galaxies, as indicated by the power-law fit parameters
in Table~\ref{tab:corr}.
Also shown in Figure~\ref{fig:xi} are the correlation functions for three,
volume-limited
samples, with luminosities centered on $M^* - 1.5$, $M^*$ and $M^* + 1.5$
(bright, medium and faint).
The power-law slopes for these samples are all consistent with $\gamma = 1.8$, 
although the correlation length $r_0$ decreases as 
expected from bright to faint luminosities.

\subsection{Galaxy-mass correlation function}

\begin{figure}
\parbox{95mm}{
\includegraphics[width=95mm]{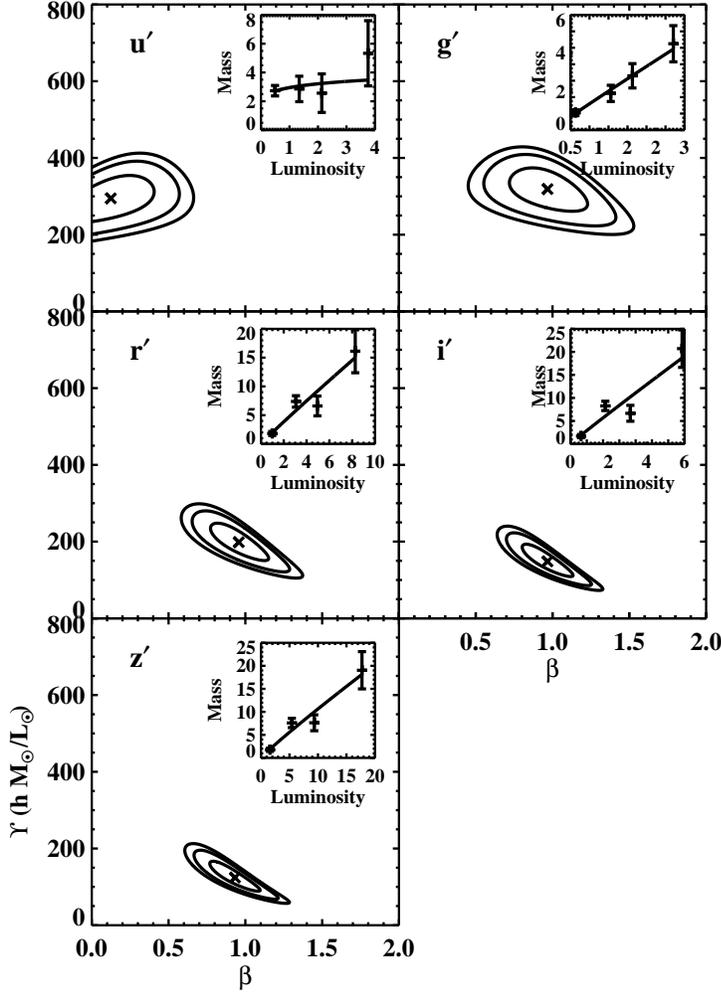}
}
\hfill
\parbox{7cm}{
\caption{Mass-luminosity relation in the five SDSS bands estimated from 
weak lensing.
The inset in each panel plots estimated mass within $260 h^{-1}$kpc ($M_{260}$)
as a function of lens luminosity.
The contours show 1, 2 and 3 sigma confidence limits on the scale factor
$\Upsilon$ and the power-law index $\beta$ in the relation
$M_{260} = \Upsilon(L/10^{10}L_\odot)^\beta$.
Note that inferred mass has only very weak dependence on $u$-band luminosity,
but in the redder survey bands $griz$, the mass-luminosity relation appears 
to be linear.
From~\cite{mckay2001}.}
\label{fig:mckay}
}
\end{figure}

So far, I have summarized recent SDSS results concerning the distribution
of {\em luminous} matter in the Universe.
Direct constraints on the {\em dark} matter distribution may be obtained
from gravitational lensing, in which the images of background sources are
distorted by the gravitational field of foreground masses.
McKay et al. \cite{mckay2001} have made weak lensing measurements of the
surface mass density contrast around foreground galaxies of known redshift.
Although the lensing signal is too weak to detect about any single lens,
by stacking together around 31,000 lens galaxies a clear lensing signal is
detected.
The galaxy-mass correlation function is well fit by a power-law of the form
$\Delta\Sigma_+ = 2.5 (r/\mbox{Mpc})^{-0.8} h M_\odot$ pc$^{-2}$,
where $M_\odot$ represents the mass of the Sun.
The strength of correlation is found to increase with the following 
properties of the lensing galaxy: late $\rightarrow$ early-type morphology,
local density and luminosity in all bands apart from $u'$.
Figure~\ref{fig:mckay} shows the relationship between inferred mass within
a $260 h^{-1}$kpc radius and luminosity in each of the survey bands.

\subsection{High-redshift quasars}

The SDSS has broken the $z=6$ redshift barrier, with the discovery of
a quasar at a redshift $z=6.28$, along with two new quasars at redshifts 
$z=5.82$ and $z=5.99$ \cite{fan2001}.
These objects were selected as $i$-dropouts: $i^* - z^* > 2.2$ and 
$z^* < 20.2$.
Contaminating L and T dwarfs were eliminated with followup near-IR 
photometry and confirming spectra were obtained with the ARC 3.5m telescope.
The SDSS has now observed a well-defined sample of four luminous quasars 
at redshift $z>5.8$.
The Eddington luminosities of these quasars are consistent with a 
central black hole of mass several times $10^9$ M$_\odot$,
and with host dark matter halos of mass $\sim 10^{13}$ M$_\odot$.
The existence of such mass concentrations at redshifts $z \approx 6$,
when the Universe was less than 1Gyr old, provides important constraints
on models of formation of massive black holes.
We expect to discover $\sim 27$  $z > 5.8$ quasars and one $z \approx 6.6$
quasar by the time the survey is complete.
Such observations will set strong constraints on cosmological models for
galaxy and quasar formation.

\section{Conclusions and Acknowledgments}

The Sloan Digital Sky Survey is now fully operational and is producing
high quality data at a prodigious rate.
We have imaged \area\ deg$^2$ of sky in five colours and have obtained
more than \nspec\ spectra.
Much exciting science has already come out of just a small fraction of
the final dataset and we look forward to many more exciting discoveries in
the coming years.

Funding for the creation and distribution of the SDSS Archive has been
provided by the Alfred P. Sloan Foundation, the Participating
Institutions, the National Aeronautics and Space Administration, the
National Science Foundation, the U.S. Department of Energy, the
Japanese Monbukagakusho, and the Max Planck Society. The SDSS Web site
is {\tt http://www.sdss.org/}.

The SDSS is managed by the Astrophysical Research Consortium (ARC) for
the Participating Institutions. The Participating Institutions are The
University of Chicago, Fermilab, the Institute for Advanced Study, the
Japan Participation Group, The Johns Hopkins University, Los Alamos
National Laboratory, the Max-Planck-Institute for Astronomy (MPIA),
the Max-Planck-Institute for Astrophysics (MPA), New Mexico State
University, Princeton University, the United States Naval Observatory,
and the University of Washington.

It is a pleasure to thank SDSS colleagues for supplying some of the figures.
I would particularly like to thank Donald York for
his careful reading of the manuscript.

{\em Jon Loveday} is a Lecturer in the Astronomy Centre, University of Sussex.
He has been interested in galaxy surveys and observational cosmology
since studying for his PhD in Astronomy at the University of Cambridge.
After spending three years in Australia, he became one of the first SDSS
participants while at Fermilab and then the University of Chicago.
He is still occasionally seen carrying a violin case.

\vfill
\end{document}